\documentclass[useAMS,usenatbib]{mn2e}

\title[Exponential profiles]
{Exponential profiles from stellar scattering off interstellar clumps and holes in dwarf galaxy discs}

\author[C. Struck and B. G. Elmegreen]
{Curtis Struck,\thanks{E-mail: curt@iastate.edu (CS);
bge@us.ibm.com (BGE)}$^1$
Bruce G. Elmegreen$^{2}$ \\
$^1$ Dept. of Physics and Astronomy, Iowa State Univ., Ames, IA 50011 USA\\
$^2$ IBM Research Division, T.J. Watson Research Center, 1101 Kitchawan Road,
Yorktown Heights, NY 10598, USA}
\usepackage{amssymb}
\usepackage{amsmath}
\usepackage{graphicx}
\usepackage{multirow}

\def\apj{{ApJ}}

\begin{document}
\date{\today}

\pagerange{\pageref{firstpage}--\pageref{lastpage}} \pubyear{0000}

\maketitle

\label{firstpage}
\begin{abstract}
Holes and clumps in the interstellar gas of dwarf irregular galaxies are
gravitational scattering centers that heat field stars and change their radial and
vertical distributions. Because the gas structures are extended and each stellar
scattering is relatively weak, the stellar orbits remain nearly circular and the
net effect accumulates slowly over time. We calculate the radial profile of
scattered stars with an idealized model and find that it approaches an equilibrium
shape that is exponential, similar to the observed shapes of galaxy discs. Our
models treat only scattering and have no bars or spiral arms, so the results apply
mostly to dwarf irregular galaxies where there are no other obvious scattering
processes. Stellar scattering by gaseous perturbations slows down when the stellar
population gets thicker than the gas layer. An accreting galaxy with a growing thin
gas layer can form multiple stellar exponential profiles from the inside-out,
preserving the remnants of each Gyr interval in a sequence of ever-lengthening and
thinning stellar subdiscs.
\end{abstract}


\section{Introduction}
\label{intro}

The origin of the exponential radial profile for surface brightness and mass surface
density in galaxy discs is not well understood \citep{vdk11}. Cosmological collapse
models get exponential discs without fine tuning
\citep{sanchez09,cooper13,aumer13,aumer13b,stinson13,martig14,herpich15,minchev15,rathaus16},
so the angular momentum distribution of the gas in the halo is one important factor
\citep{mestel63,freeman70}. This may explain why the exponential profile is established
early on, out to redshift $z=5.8$ in \cite{fathi12} and in clumpy discs at $z\sim2$
\citep{elmegreen05}, and why it is also present in extended outer gas discs today, where
there has been little star formation \citep{bigiel12,wang14}.

Other processes move stars radially in the disc after they form. Wet major mergers
in the simulations by \cite{athan16} formed a Type I disc in the reaccreted
remnants \citep[Type I is a single, continuous exponential profile,
see][]{freeman70}. Mergers in \cite{borlaff14} selected from a hydrodynamic N-body
database to resemble S0s tended to have Type III exponentials
\citep[i.e., profiles with two exponential segments, the outer one shallower than the inner one;][]{pohlen06}.
Stars also scatter outward with bar torques
\citep{hohl71,debattista06} forming type II (a double exponential with a steeper
outer segment) or Type III exponentials \citep{head15}, with breaks between the two
components at bar or bar+spiral resonances \citep{munozmateos13,laine14}. Stars
churn back and forth through corotation in spiral arms \citep{se02}, causing radial
migration when there are recurrent transient spirals
\citep{roskar08,minchev12,donghia13,veraciro14,gr15,be15} and mixing stars with
different elemental abundances \citep{roskar08} in agreement with observations
\citep{edvardsson13}. Satellite galaxies \citep{gr15b}, and giant gaseous clumps in
the disc \citep{bournaud07} also move stars radially.

The question we wish to address here is why these relocated stars settle into an
exponential form, or a piecewise exponential form for the Types II and III. For
example, it seems remarkable that galaxies with U-shaped color
\citep{bakos08,yoachim12} or age \citep{roediger12,ruiz16} profiles can have a mass
profile that is a single exponential even if the light profile is Type II
\citep{zheng15}. How does the disc mass know it should have an exponential form when
the two parts form in different ways, the inner part by star formation with spiral and
bar torques, and the outer part by radial migration? Why should both parts of the
broken profiles in Types II and III be exponential, regardless of whether the stars
were put there by bars, spirals, interactions, or star formation? There is something
fundamental about the exponential shape in a galaxy disc. In \citet[][hereafter Paper
I]{es13}, we considered the driver to be random scattering itself, regardless of the
source. A theoretical stochastic model for scattering with this property was discussed
in \cite{elmegreen16}.

A good laboratory for studying these questions is a dwarf Irregular (dIrr) galaxy,
which has no spirals and often no bar or obvious companions \citep{hunter06}, and yet
it has an exponential or piecewise exponential radial profile in optical through
infrared light and in mass \citep{hunter11,herrmann13,herrmann16}.   If stellar
scattering is important for establishing or maintaining this profile, then the
scattering masses cannot be spirals or bars. Their stellar discs are usually smooth.
Even the stellar clusters and OB associations that make them look irregular in
H$\alpha$ and blue light have little mass compared to the underlying discs
\citep{zhang12}.

Stellar scattering off of gas irregularities in dIrr galaxies may be a good
alternative to scattering off of spirals and bars. The surface densities of many
dIrrs are dominated by gas, especially in their outer parts. For example, figures
24-27 in \cite{leroy08} show the radial profiles of surface mass density for gas and
stars in 4 dIrr galaxies. The gas dominates the stars by a factor from 1.4 in the inner
parts to over 8 in the outer parts in DDO 154, it dominates by a factor of 5 throughout
Ho I, and by a factor up to $\sim10$ in the main disks and outer parts of the other two
dIrrs. Also, the middle column of figure 8 and table 2 in \cite{eh15} compare gas and
stellar surface density profiles in 20 local dIrr galaxies in the LITTLE THINGS
survey. The average surface density for stars ranges from $4.87\;M_\odot$ pc$^{-2}$ to
$0.134\;M_\odot$ pc$^{-2}$ between the center and 4 disk scale lengths in these 20
galaxies, while the average surface density for gas ranges from $9.77\;M_\odot$
pc$^{-2}$ to $1.589\;M_\odot$ pc$^{-2}$ over the same radial range. Thus the gas ranges
between a factor of 2 higher surface density than the stars in the inner part to a
factor of 10 higher in the outer part for these dIrrs. Similarly, in a survey of
148 isolated dwarf galaxies in the stellar mass range from $10^7\;M_\odot$ to
$10^{9.5}\;M_\odot$ \citep[the upper limit is approximately the mass of the Large
Magellanic Cloud;][]{james11}, the median atomic gas fraction among baryons is
$0.81\pm0.31$ and the gas fraction inversely correlates with stellar surface density
\citep{bradford15}. Gas-rich dwarfs like this are distinct from dwarf Spheroidals.
Local group dwarfs in the stellar mass range $10^6-10^7\;M_\odot$ are gas poor inside
the virial radii of the Milky Way and M31, presumably because of environmental effects
\citep{spekkens14,weisz14}.

Extensive observations show that the gas is highly irregular in gas-rich dIrrs.
There are large clouds of atomic gas \citep{hunter01,kim07} and numerous shells and
holes that result from star formation feedback
\citep{yang93,kim99,loz09,egorov10,warren11,bagetakos11,kaczmarek12,pokhrel16,egorov16}.
These structures extend for tens to hundreds of parsecs.  The largest may contain a
total mass or mass displacement amounting to one percent of the galaxy stellar mass.
For example, the dIrr galaxy IC 2574 has a dynamical mass of $8.3\times10^9\;M_\odot$
and contains $\sim50$ HI holes with the largest displacing $1.5\times10^7\;M_\odot$ of
gas and 7 others displacing over $10^6\;M_\odot$ of gas \citep{walter99}. Similarly, Ho
II has a dynamical mass of $2\times10^9\;M_\odot$ and a largest hole displacement of
$1.5\times10^7\;M_\odot$ in gas;  11 others displace over $10^6\;M_\odot$
\citep{puche92}.

Interstellar gas irregularities of this magnitude should deflect the orbits of field
stars \citep{spitzer53}.  The origin of the irregularities themselves is not important.
Giant clouds could form in the compressed regions between shells and supershells, for
example, \citep[e.g.,][]{chernin95,yamaguchi01,inutsuka15}, especially if the Toomre
$Q$ parameter is too large for gravitational instabilities to be active \citep[as
observed by][]{eh15}. If a particular irregularity lasts for a shear time
\citep{bagetakos11}, then that would be long enough to have a significant cumulative
effect on stars for some distance away. This distance can be determined as follows.
Consider stars a distance $d$ away from a cloud, measured in the plane, and moving at a
speed that is the stellar velocity dispersion, $\sigma$. If $\Omega$ is the angular
rotation rate of the galaxy and $\Omega^\prime$ is its radial derivative, then the time
for shear to move parts of the cloud further from the star than the distance $d$ is
$1/(d\Omega^\prime)$. The time for the star to move past the cloud is $d/\sigma$. For
significant scattering by that one cloud, we need the first time to exceed the second,
or $d<(\sigma/\Omega^\prime)^{0.5}$. For a rotation curve $V(R)\propto R^\alpha$, this
limit becomes $d/R<(\sigma/\left[1-\alpha\right]V)^{0.5}$. Rotation curves in dIrr
galaxies are slowly rising, so $\alpha$ is significantly larger than 0, e.g.,
$\sim0.5$. Also, the rotation speeds are slow so $\sigma/V\sim0.3$ or more. Then $d/R$
can be large, like 0.8 or 1, which means that a significant number of stars have time
to react to the force of the cloud before it shears away.

{The transient nature of interstellar clouds and holes does not matter much for
stellar scattering anyway. As long as clouds and holes are present on a
steady-state basis, with new structures replacing old structures regularly, there
will always be the types of perturbations that we model here, even if the
individual forcings change with time. If one cloud starts to scatter a particular
star and then a younger cloud takes over after a short time, the star is still
scattered. The most important property of scattering is that it changes stellar
orbits from approximately circular at birth, owing to the low velocity dispersion
of the gas, to increasingly eccentric after the star random walks for a while among
numerous scattering centers. Also, because the cloud and hole perturbation masses
are much larger than a star's mass, each scattering event approximately conserves
stellar orbital energy. The result is a net loss of angular momentum for an
increasingly eccentric orbit at constant energy. Thus scattering like this has a
net bias toward smaller radius, and then the exponential radial profiles follows
mathematically as an equilibrium solution to the stochastic process
\citep{elmegreen16}.}

Large clouds and holes in dIrrs are soft mass perturbations that scatter stars by
relatively small amounts with each interaction. They are like flocculent spiral
arms in this respect \citep{veraciro14}. Still, their cumulative effect can be
important for the long-term rearrangement of orbits (Paper I). This situation
differs from the case of young galaxies where interstellar clumps are more massive
compared to the galaxy, and stellar scattering is faster \citep{bournaud07}. It
also differs from stellar scattering in a time-changing central potential
\citep{governato12,el15}, which is rapid too. In both cases the resulting orbits
are eccentric, as observed today in the Milky Way thick disc \citep{liu12}, which
formed at these early times \citep{bournaud09,bird13,robin14,comeron14,aumer16}.

Here we show that large, low-density perturbations with a range of masses
comparable to what is observed for interstellar clumps and holes in dIrr galaxies
can make an exponential profile from a uniform initial profile in only $\sim1$ Gyr.
This initial profile is arbitrary, selected as an extreme case of a non-exponential
shape only to show the speed of the profile adjustment. If the distributions of
halo angular momentum and disc torques give a different initial profile for stars,
or if a disruptive event changes the profile, then cloud and hole scattering will
return the disc to an exponential.

Stellar scattering off interstellar clouds has long been viewed as a mechanism to
increase the stellar velocity dispersion \citep{spitzer53,la84}. Observations of
the dIrr galaxy WLM \citep{leaman12} and theoretical considerations
\citep[e.g.,][]{gust16,aumer16} support this model. These studies did not consider
the influence of cloud-star scattering on the stellar radial profile, however, and
that is the purpose of the present work. We also illustrate that interstellar holes
are as effective in scattering stars as interstellar clouds. In a real dwarf
Irregular galaxy, both holes and clouds should act together to increase the stellar
velocity dispersion and form or maintain the exponential radial profile.

\section{Models}
\label{sect:models}

We wish to isolate the physical process of greatest interest here, which is stellar
scattering from mass structures in the disc of a galaxy. We do not want these
processes to be confused with additional things that happen in galaxies, such
scattering from bars and spirals. This isolation of pure scattering is a great
simplification -- not unlike the simplification made for analytical models of
star-GMC scattering \citep{la84} except that here we can study the redistribution
of stars in radius too.  The approximation is also reasonable for dIrr galaxies
because they have no spirals.

The model is based on a numerical integration of the motion of approximately 15,600
test particle `stars' using the MATLAB Runge-Kutta routine `ode23.' These particles
are initially placed in circular orbits in a fixed gravitational potential and
allowed to interact gravitationally with some finite number of scattering objects
(clouds or holes).

We extend our previous two-dimensional models in several ways. First, the new
models are computed in three dimensions to account for saturation effects when the
disc gets thick (Sec. \ref{three}). Second, we use diffuse clumps in one case and
holes in another case. Both clumps and holes perturb the orbits of stars. Third, we
use three-part mass and size spectra to represent these perturbations, following
observations in \cite{kim07}, \cite{oey97}, \cite{bagetakos11} and others.  These
spectra seem to have a universal slope in galaxies. Analytical models show that the
stellar heating rate from cloud scattering depends on the product of the cloud
surface density and the cloud mass \citep{la84}. For a power law cloud mass
spectrum, the rate depends on the integral over this spectrum. If the slope of the
mass spectrum is shallower than $-1$ on a log-log plot, then this integral is
dominated by the mass of the largest cloud, and the spectral slope enters only
weakly as a coefficient. The same argument applies to star-hole scattering. This is
the case here, so the results do not depend much on the mass spectra.

We begin with the parameters used to scale the (dimensionless) calculations. In
Paper I we used two sets of scaling parameters, one for a typical spiral disc and
another for a dwarf galaxy disc. Here we are most interested in the latter case and
adopt the scaling factors used for it. Specifically, these include a gravitational
or halo scale length $H = 0.5$ kpc, and a corresponding velocity of $V_{\rm H} =
50$ km s$^{-1}$. Then we have a time scaling of $T = 9.8$ Myr, a mass scale of
$M_{\rm H} = 2.9 \times 10^8\;M_{\odot}$ and a rotation period at $R = H$ of $61$
Myr.

We use a (dark halo) gravitational acceleration of the form,

\begin{equation}
\label{eqa}
g(r) = \frac{-GM_H}{H^2}
\left(\frac{r + 0.2}{H} \right) ^{-0.4},
\end{equation}

\noindent where the 0.2 factor is an arbitrary softening constant, and $r$ is the
central radius in three dimensions. In this potential the disc has a circular
velocity that increases gradually as $R^{0.3}$, where $R$ is the projected radius
within the disc, and a gravitational mass within a radius $r$ of $M(r) = M_H
(r/H)^{1.6}$. This slowly rising rotation curve is comparable to that observed in
dIrr galaxies \citep{oh15}. We choose an initial maximum disc size of 9 units,
which equals $4.5$ kpc in the adopted scaling and has a halo mass enclosed within
that radius of $1.9 \times 10^{10}\; M_\odot$. We estimate the baryonic mass as
about one-tenth of this, and the gas mass as about 40\% of that, or $7.6 \times
10^8\; M_\odot$. These values are not based on any specific model, but are
consistent with the observations of IC 2574 \citep{walter99}.

An additional force is added in the vertical ($z$) direction to roughly represent the
local disc gravity. The adopted form of this acceleration is linear in $z$, and
with a magnitude that allows a reasonably thick disc, as observed in dIrrs
\citep{eh15}. Specifically,

\begin{equation}
\label{eqaa}
g_z(r) = \frac{-0.3GM_H}{H^2}
\left( \frac{z}{H} \right).
\end{equation}

\noindent Other forms of vertical acceleration were tested, including a constant and one
that scales as $z/R$. These altered the disc thickness, but did not significantly change
other results described below.

As in \citet{es13} the perturbing clumps or holes are initially distributed
uniformly in radius (giving a $1/r$ surface density profile) and at random azimuths
on circular orbits. Clumps or holes of all different sizes (see below) are
distributed in the same manner. Realistically, the larger clumps and holes might be
more confined to the central regions, in which case we are over-estimating, or
speeding up, scattering in the outer regions. The clumps follow circular orbits at
their initial radii for the whole of the simulation. This circularity is another
approximation because real clumps and holes should grow and disperse over time,
possibly also changing their positions in the disc.  We showed above that shear is
too slow to affect the results as long as the number density of the irregularities
is about constant in time.  Chance alignments of the scattering centers generate
small stellar wakes.

The fiducial clouds model uses a clump mass spectrum with a slope of $-0.5$ in a
log-log distribution. This is similar to the observed cloud mass spectrum in a the
Milky Way \citep{heyer01} and LMC \citep{kim07}. Then, with $4$ of the most massive
clouds, each of mass $0.105$ units $= 2.9 \times 10^7\; M_\odot$, we use $9$
intermediate-mass clouds with $5.8 \times 10^6\; M_\odot$ each, and $59$ low-mass
clouds with $2.0 \times 10^5\; M_\odot$ each.  This adds up to a total of $1.8
\times 10^8 M_\odot$; any additional gas is assumed to be in a smooth form that
does not scatter stars.

The corresponding cloud radii are: $0.8$ units $= 400$ pc, 200 pc and 100 pc,
respectively, which implies that the size spectrum has an average slope of $-1.9$
on a log-log plot. The cloud radius is simulated as a smoothing length for the
equation of force between the cloud and the star particles. The star particles do
not interact with each other, so there is no smoothing length needed for star-star
gravity. Also because the stars are non-interacting, their number in the simulation
does not matter for the result except to make a smooth enough distribution that the
disc evolution can be seen.

For the holes models, we assume hole sizes and masses like the clump sizes and
masses of the previous paragraph, but with negative mass.   For comparison,
\cite{oey97} observed a hole size spectrum for the Small Magellanic Cloud with a
slope on a log-log plot of $-1.8\pm0.4$, while \cite{bagetakos11} observed a slope
of $-1.9$, in agreement with our value. In order to estimate the mass that can be
evacuated from specified holes we need a rough model of the gas disc.  We assume
that the gas surface density profile also has an approximately exponential form,

\begin{equation}
\label{eqb}
\Sigma = \Sigma_1 e^{-\frac{r}{a}},
\end{equation}

\noindent where subscript `1' denotes the inner radius, e.g., $r_1 = H$. Then,

\begin{equation}
\label{eqc}
M_g(r) = \int^{r}_{r_1} dr \left( 2\pi r \Sigma \right),
\end{equation}

\noindent and,

\begin{equation}
\label{eqd}
\Sigma_1= \frac{M_g(r_2)}{2\pi a^2}
\left[ \left( \frac{1 + r_1}{a} \right) e^{-\frac{r_1}{a}}
- \left( \frac{1 + r_2}{a} \right) e^{-\frac{r_2}{a}} \right]^{-1} ,
\end{equation}

\noindent where subscript `2' denotes the outer radius, and $M_g(r_2)$ is the total gas
mass. With the scalings adopted above, and assuming that $a \simeq 2.0$ units ($1.0$
kpc), this latter equation yields $\Sigma(r_1) \simeq 39\;M_\odot$ pc$^{-2} \simeq 4.3
\times 10^{22}$ cm$^{-2}$. This column density declines to about $\Sigma(r_2) \simeq 7.9
\times 10^{20}$ cm$^{-2}$ at the outer edge of the disc. If we assume an inner disc
thickness of about $200$ pc, then the volume density corresponding to the column density
there is about $n(r_1) \simeq 70 $ cm$^{-3}$.

There is a lower limit to the negative mass of the hole particles because the total
density cannot go negative. The sizes and masses in the fiducial clump are close to
this limit. For example, if we want a hole in the inner disc with an equivalent
mass of $-2.9 \times 10^7\; M_\odot$, then using the density estimates of the
previous paragraph, the size of the hole must be about $486$ pc, which is
comparable to the $400$ pc assumed. In the other two the appropriate minimum hole
sizes for the given masses are: $218$ pc and 40 pc. These minima should be larger
by a factor of $\sim 7$ in the outer disc where the density is lower.

\section{Model Results}

\subsection{Formation of the Exponential}

\begin{figure}
\centerline{
\includegraphics[scale=0.42]{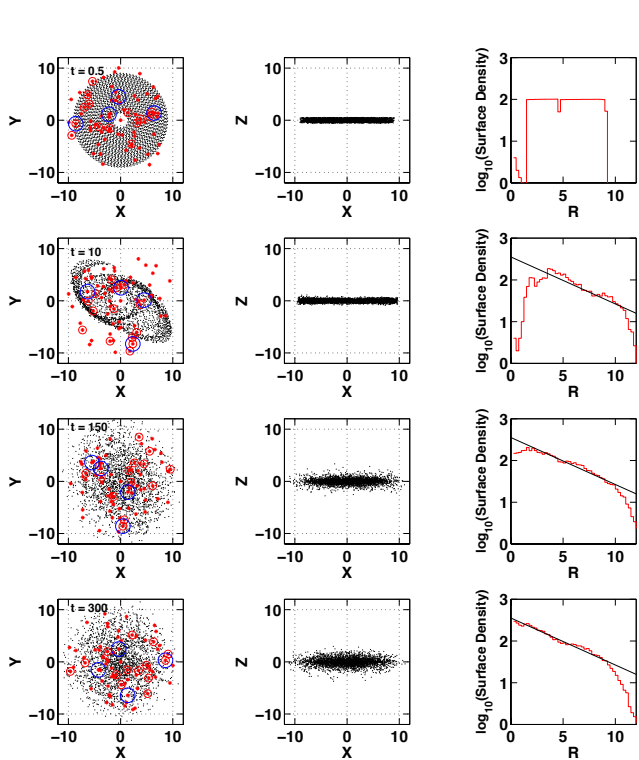}}
\caption{Four snapshots showing face-on and side views, and the development of the
exponential profile in the fiducial soft clump model. Times given in simulation
units of 9.8 Myr for a typical dIrr galaxy. In the first column, the red asterisks
represent the smallest mass clumps, the red circles the intermediate mass clumps,
and the blue circles the most massive clumps, in each case not to scale. Every
seventh simulation particle plotted. Note that the initial central hole fills in at
later times, so the exponential form grows inward and outward.}
\label{fig:foursnap}\end{figure}

\begin{figure}
\centerline{
\includegraphics[scale=0.37]{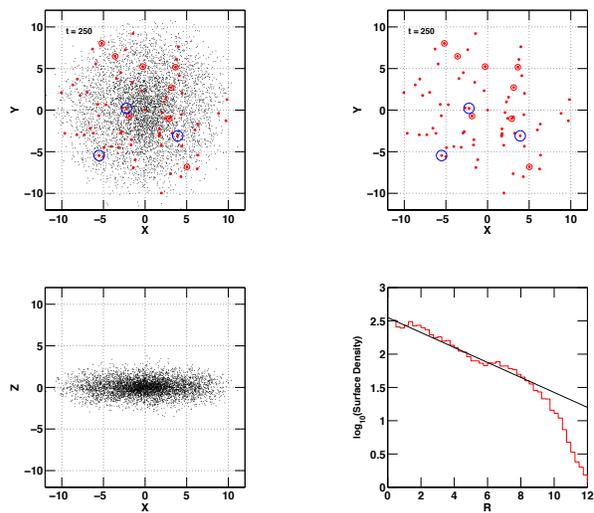}}
\caption{Snapshot and profile graph at $t = 250\ units$ in the holes model, with a panel of
the perturbs only in the upper right. The solid line is the same as that in the
last panels of Fig. 1, showing near identity of the profile in this model. The red
asterisks and red and blue circles represent holes of different size (not to
scale), like the clumps shown in Fig. 1. The surface density falls off by about a
factor of 10 from the inner part to the outer downturn.}
\label{fig:comp}\end{figure}

\begin{figure}
\centerline{
\includegraphics[scale=0.48]{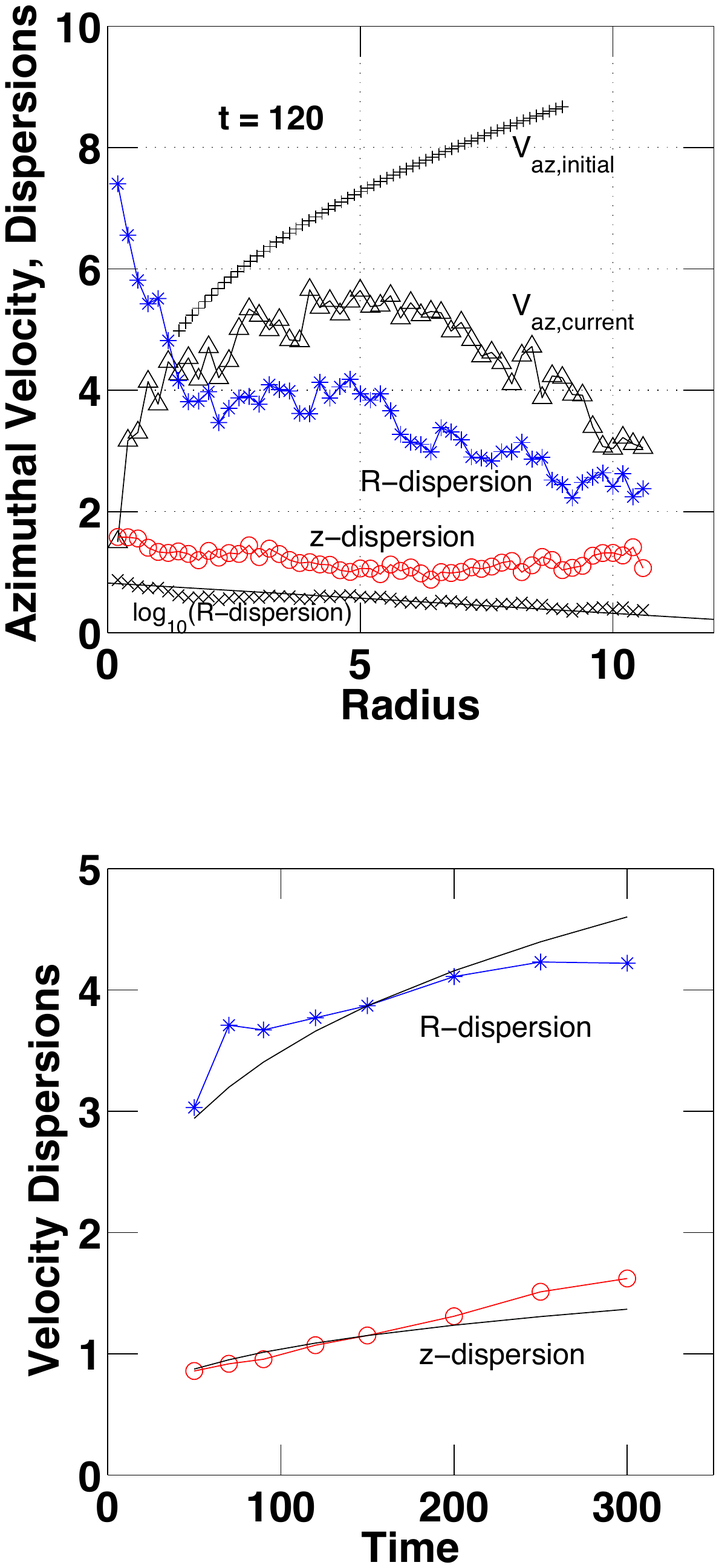}}
\caption{Velocity profiles in the clump model of Fig. 1 are shown in the top panel. The
black plus signs denote the initial rotation curve given by the gravitational
potential with zero velocity dispersion. The black triangles show the
stellar rotation curve from actually stellar motions
at a time of $120$ units ($\sim1.2$ Gyr). Each point is the
average azimuthal velocity in a radial bin of width $0.25\ units$ ($\sim125\ pc$).
The blue asterisks and red open circles show the radial and vertical velocity
dispersion profile at that time. The black x symbols show the logarithm of the
radial dispersion, with a straight line drawn to show that this profile is
approximately exponential. The bottom panel shows the radial (blue asterisks) and
$z$ (red circles) velocity dispersions, averaged over the radial interval (4.0,
6.0) as function of time. The black curves are $t^{1/4}$ curves, proposed by
\citet{la84}, and here normalized to the numerical results at $t = 150$. The radial
dispersion appears to saturate at late times, while the z-dispersion grows nearly
linearly. } \label{fig:vels}\end{figure}

The results are summarized in Figures \ref{fig:foursnap} to \ref{fig:vels}. Figure
\ref{fig:foursnap} shows that scattering from large gravitationally soft clumps can
convert an initially flat stellar density profile into an exponential profile in a
relatively short time. The times given in the left-hand panels are in units of 9.8 Myr.
The straight lines in the right-hand panels are all the same and meant to guide the
eye.

Figure \ref{fig:comp} comes from a model with holes, rather than clumps, and shows that
holes in an otherwise smooth disc form exponential stellar profiles in essentially the
same way as clumps. Real discs contain clumps, shells, and holes on a variety of
scales, but regardless of the structural details, our results suggest that an
exponential stellar profile is a general outcome. This confirms the primary result of
\citet{es13}, and generalizes it to the cases of softer scatterers, a spectrum of
scatterers, and to scattering in three dimensions.

The top panel of Figure \ref{fig:vels} shows the initial rotation curve of the clump
model (Fig. \ref{fig:foursnap}), using the gravitational potential with no random
motions, and it shows the actual stellar azimuthal speed and velocity
dispersion profiles at a time of $\sim1.2$ Gyr. By this time, the exponential surface
density profile is well developed, the stellar rotation curve rises in the centre and
then falls in the outer disc where the scattered stars are, and the velocity dispersion
profile falls with radius.

The radial component of the dispersion can be approximately fit by an exponential, as
in \citet{martin13}. This fit is shown by the x-marks at the bottom of the figure,
which are plotted on a log-linear scale.  The eccentricity of the stellar orbits
depends on the softness of the scattering structures. For the soft scatterers here, the
exponential profile, the highly eccentric orbits and the resulting velocity dispersion
all build up more slowly than in \citet{es13}.

The bottom panel of Figure \ref{fig:vels} shows the time evolution of the two
dispersions. Specifically, the dispersions were averaged over the radial interval of
(4.0, 6.0) at selected times. Early times, dominated by transient waves, were not
included. Both dispersions are negligible at the start of the run. The radial
dispersion grows rapidly and evidently saturates at late times. The vertical ($z$)
dispersion grows more slowly, but steadily. As shown in Figure \ref{fig:vels} both
dispersions can be approximated by a $t^{1/4}$ function, which was derived analytically
for molecular cloud scattering by \citet{la84}. However, the fit is not particularly
good, and the z-dispersion growth is better fit by a straight line.

\begin{figure}
\centerline{
\includegraphics[scale=0.35]{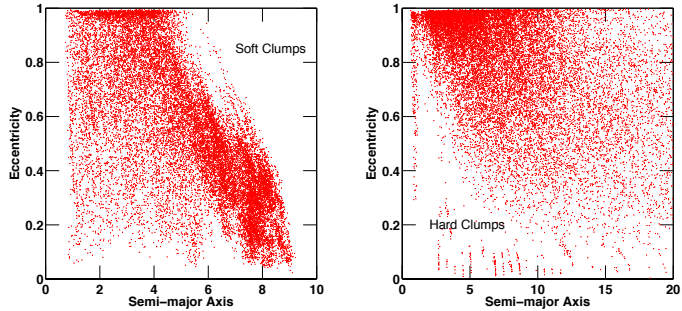}}
\caption{Two views of the distributions of the estimated semi-major axis and eccentricity
values for test particle orbits in scattering simulations. Left panel: from a
simulation run with essentially the same parameter values as the runs shown in the
previous figures, except with more particles (20,187), and run in two dimensions
for consistency with the right panel. Right panel: data taken from the rising
rotation curve run of \citet{es13}, which used compact scattering centres
(clumps).} \label{fig:ecc}\end{figure}

\subsection{Eccentricities of Stellar Orbits}

Figure \ref{fig:ecc} shows the orbital eccentricity distributions as a function of
semi-major axis size for the cases of hard and soft clump scattering. (Eccentricity is
defined here as the difference between the maximum and minimum orbital radii divided by
their sum.) The hard clump case is taken from the rising rotation curve model of
\citet{es13}. The soft clump case has parameters like those of the models discussed
above, except it is run in two dimensions for consistency with the hard clump model,
and has the same number of test particles as that model. In both cases the orbital
parameters were estimated by using the average of several of the largest and smallest
radii reached by each test particle over a time of 30 units from a starting time of 150
units. This procedure is a compromise because it averages over less than a whole orbit
for the particles with the largest orbital radii, but can average over orbital changes
for particles with the smallest orbital radii. Thus, the derived orbital parameters are
only estimates.

Figure \ref{fig:ecc} shows important differences between the soft and hard scattering
cases. The most apparent is that the soft scattering case has a large population of
particles with small to moderate eccentricities, while the hard scattering case has
many fewer of these particles. In the soft scattering case these moderate eccentricity
particles are found at all radii, although they are concentrated at the larger radii.
This is true even from early times in that run. Note that the x-axis in the hard
scattering case has twice the range of the soft-scattering case. In the hard scattering
case, there are very few particles with semi-major axes less than $a = 8.0$ and
eccentricities less than $0.4$. Thus, the soft-scattering models greatly decrease the
number of high eccentricity orbits relative to the hard scattering models of
\citet{es13}.

In both cases there is a concentration of particles on orbits with small radii and large
eccentricities. In part, this may be an artifact of the initial conditions, where massive
clumps are simply placed among the test particles, rather than having a chance to scatter
more mildly while growing. However, a part of this population of eccentric orbits would
likely develop in any case. These orbits are strongly scattered, and in reality that
likely result in a component out of the disc. In that case, these would be viewed as bulge
or halo stars.

If, on these grounds, we exclude the particles with high eccentricities (e.g., $e >
0.8$), then the eccentricity distribution in the soft scattering solution is
qualitatively similar to N-body/hydrodynamic model results shown in  Figure 9 of
\citet{in14}.  These latter simulations model the evolution of clumpy discs in
particular, and so provide a comparison to our simple scattering models, though
comparisons are complicated by the fact that they also include accretion from a cooling
halo. Specifically, the \citet{in14} models produce thick discs after a time of several
Gyr, like the model of Figure \ref{fig:foursnap}.

The dependence of the stellar velocity dispersion and orbital eccentricity on the
hardness of the scattering centers suggests a possible variation in the relative
thicknesses of dIrr disks and in the ratios of the rotation speeds to the dispersions.
The models in figure \ref{fig:vels} with soft scattering centers have azimuthal
velocities that are $\sim50$\% larger than the radial dispersions and $\sim4\times$
larger than the vertical dispersions. Figure \ref{fig:ecc} suggests that these ratios
could be smaller with harder scattering centers. Dwarf Spheroidal galaxies have small
velocity ratios, as mentioned above, but they are more evolved than anything considered
here. The small local dIrr WLM has a rotation-to-dispersion ratio for stars of order
unity, which the observers, \cite{leaman12}, also suggest is a result of star-cloud
scattering.

\subsection{Three Phases of Evolution}
\label{three}

Figure \ref{fig:foursnap} shows three evolutionary phases in our models. In the first,
the initially circular orbits are strongly disrupted by scattering and form transient
waves. The organization of these waves depends on the accidental coherent forcing from
the initial distribution of the large clumps. They generally decay in the second phase,
though they can be re-generated by a fortuitous conjunction of scattering centres.
Examination of their early evolution suggests that they originate in the strong
backward pull on particles orbiting ahead of the most massive clumps (or forward pull
in the hole model). If their self-gravity were included these waves would serve as
additional soft-scattering structures \citep{to91} to amplify the gravity of the
original ISM irregularities and shorten the evolution timescale.

In the second phase, which occurs between 50 and 150 time units (500 to 1500 Myr), the
transient waves diminish while the surface density, velocity dispersion, and disc
thickness evolve slowly. Early in this phase, the surface profile is approximately a
double exponential with a break near the initial disc radius, but by the end of phase
2, the outer part comes up and a single exponential appears out to a disc edge.
Throughout this phase the kinematics are dominated by rotation, but the radial and
vertical velocity dispersions grow to the levels shown in Figure \ref{fig:vels}.

In the third phase, the development of the exponential slows. The radial dispersion and
azimuthal drift of the rotation curve have increased to the point where the dispersion
dominates the rotation, and these quantities continue to increase after that.  Then the
disc becomes more like a discy elliptical, dwarf Spheroidal, or E/S0 galaxy, with hot
kinematics like those observed in the Sauron \citep{em04}, ATLAS$^{3D}$ \citep{kr13} and
SLUGGS \citep{ar14} surveys, or the dispersion-dominated dwarf discs of \citet{wh15}. In
fact, the feedback processes that generate large holes and clumps in our simulations may
be reduced before this dispersion-dominated phase appears if the gas is consumed by star
formation or cleared out by feedback or ram pressure stripping.

The result that stars eventually scatter to a thicker disc is consistent with much
previous work \citep[e.g.,][]{wielen85} and with the observation that the rotation speed
of stars is lower off the midplane \citep{comeron15}. Because of dissipation, gas should
have more circular orbits than scattered stars and faster rotation speeds in the
potential well of the galaxy. This should be the case especially in the outer regions if
scattering makes the stellar orbits eccentric.

Comparison of the lower two rows of Figure \ref{fig:foursnap} shows that the profile
evolution has largely saturated by this time, evolving very slowly. The growth of the
disc thickness has also slowed (as occurs in the case of scattering by giant molecular
clouds, see e.g., \citealt{la84}). Two scattering processes affect this result. The
first is that, as noted in \citet{es13} and confirmed in additional simulations, the
stellar profile tends to grow outward in flat or falling rotation curve potentials, but
to be pulled inward in near solid-body potentials. In both cases these tendencies slow
after a couple Gyr. In intermediate potentials like that used here, the change of the
radial extent and exponential scale length is small at all times after the exponential
form is established. Secondly, in three-dimensional models, as stars are scattered and
spend more time away from the disc plane, the scattering interactions weaken. Depending
on these processes, and others like wave-induced migration, continued accretion, or
galaxy interactions, the evolution of single-age populations may be complex.

\section{Multiple Superposed Exponentials with Different Ages}

Modern galaxies appear to consist of several near-exponential profiles that are
nested together and superposed to make the overall mass profile. Each mono-age
subsystem of stars, observed within an age range of about one or a few Gyr, may
have its own profile, with a trend toward longer scale lengths and smaller scale
heights over time, i.e., for younger and younger populations. This trend has been
found in numerical simulations of Milky Way type galaxies
\citep{sanchez09,stinson13,bird13,martig14,minchev15}. Mono-abundance populations
have been observed with a similar evolution in the Milky Way \citep{bovy12,bovy16},
although mono-abundance does not necessarily translate directly into mono-age,
e.g., see \citet{minchev16}. Moreover, the Milky Way is not a dwarf galaxy, the
focus of this paper, and there are no comparable observations of dwarf discs.
However, \citet{zhang12} did find that the redder components in a sample of dwarf
discs are more extensive than young, blue components. Although not discussed in
that work, this could result from scattering as in the models above.

We expect this type of multi-component structure to result from stellar scattering
off interstellar structures because, as shown in Section \ref{three}, the evolution
of the disc slows down when it gets thick. The entire process of going from
circular orbits to saturated high-dispersion orbits takes only a few Gyr in our
models. Thus each new generation of stars formed in newly accreted disc gas has
about a Gyr to heat up, during which time it will make its own equilibrium
exponential profile starting from the initial extent of the star-forming region.
Here we show that stellar scattering off of interstellar irregularities does not
mix these different-age populations together and make a single exponential. Rather,
each Gyr period of scattering thickens whatever disk there is at that time, and
gives it an exponential profile with a scale length proportional to the current
disk size.

The roughly Gyr timescale comes also from the analytic theory in \cite{la84}. Consider
equations (54) and (57) in that paper, where characteristic timescales are given for the
cases where the stellar scale height is less than and greater than the cloud scale
height, respectively. In the first case, the basic timescale for stellar heating is
$t_1=3\sigma^3/(3D_3)$ where $\sigma$ is the initial stellar velocity dispersion and
$D_3$ is from equation (53) using evaluations from Table 2. In the second case,
$t_2=0.5\sigma^4/D_4$ where $D_4$ is from equation (56). Using $N_{\rm c}M_{\rm
c}=5\;M_\odot$ pc$^{-2}$ for cloud surface number density $N_c$ and characteristic mass
$M_c=10^6$, using $\sigma=10$ km s$^{-1}$ for initial stellar velocity dispersion, using
$h_{\rm c}=200$ pc for the cloud scale height, and using $\ln\Lambda=4.6$, we get
$t_1\sim0.6$ Gyr and $t_2\sim1$ Gyr.

\begin{figure}
\centerline{
\includegraphics[scale=0.45]{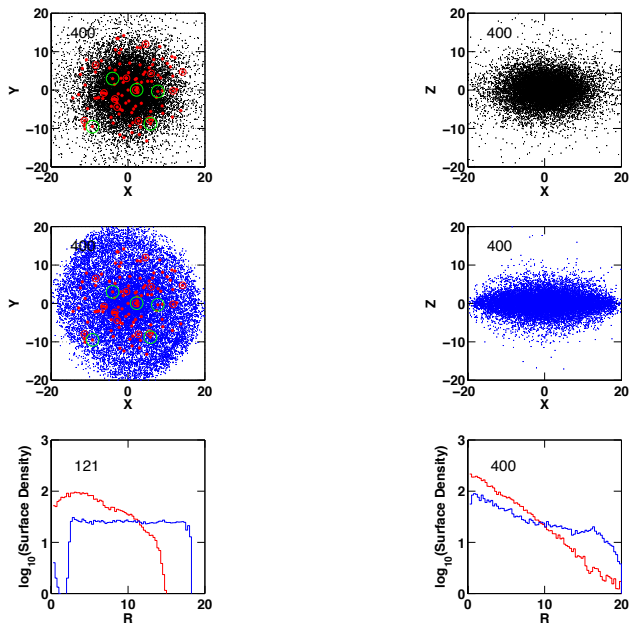}}
\caption{Model with two stellar generations  The panels of the top row show the distribution of the first generation of stars at a time of 400 units (3.9Gyr) after its formation. The left panel shows the disc plane (x-y), and the right panel a perpendicular view (x-z). The panels of the second row show the same views of the second generation at the same time. This population was introduced into the model at time of 120 units (1.2 Gyr). The bottom panel shows the surface density profiles of the two populations shortly after the younger one is introduced. The peaked red curve is the old population; the flat blue curve is the younger one. The lower right panel shows the profiles of the populations at the later time. The younger population is still flatter than the older one.}
\label{fig:2pops}\end{figure}

To compare mono-age population discs with the results of this study, we carried out an
extension of the scattering models described above to the case of two generations of
stars. This model was initiated with the same stellar population and scattering centres
as the run shown in Figure \ref{fig:foursnap}, but at a time of 120 units (1.2 Gyr).
Added to this was a second (young) stellar population with equal numbers of stars
(31,288 total). The second population was also assumed to have a flat initial density
profile (see Fig. \ref{fig:2pops}) to facilitate the study of how it evolved relative
to the older population. This second population was assumed to extend out to twice the
initial radius of the older population (compare the first row of Figure
\ref{fig:foursnap} to the young star density profile in the first row of Figure
\ref{fig:2pops}). In addition to the second generation of stars, 40 more scattering
centres were added at larger radii than the previous distribution (with the same mass
spectrum and radial profile) to enable scattering in the outer disc. These
inhomogeneities would be expected to result from cloud formation and feedback with the
new stellar generation. Modeling these inhomogeneities in the same way as for the first
generation of stars may not be realistic, but the stellar profile evolution does not
depend much on the nature of the scattering centres.

The other graphs in Figure \ref{fig:2pops} show the stellar and scattering centre
distributions at a later time of 400 units (3.9 Gyr). Both old and new stellar discs
evolve over this time interval by smoothing and expanding in the vertical direction.
The profiles of both generations continue to evolve their exponential forms. Even by
this later time, however, the two exponential profiles are not the same. The different
vertical distributions suggest different vertical velocity dispersions as well.
Generally, we would expect different mean metallicities in these two populations, so
they should be observationally distinguishable by metallicity and age despite their
overlap in radius (see the simulations of \citealt{lo11}).

We note that if, as in this model, the older population dominates in the inner
regions, while the younger does in the outer regions, then that would give rise to
a color gradient. This effect is the result of the shallow initial profiles here,
and the steady evolution to a steeper profile for a longer time in the old
population. Subject to the initial conditions, this provides a possible explanation
for the observed distributions of mono-age populations. That is, each new
population evolves into an exponential profile on a timescale of a Gyr because of
scattering off of interstellar clouds and holes, while all of the older populations
continue to evolve more slowly in the presence of the same scattering sites. In
reality, other processes, including misaligned cosmological infall, can complicate
the picture.

\section{Conclusions}

Scattering from gaseous structures like clumps and holes in a dIrr galaxy disc produce
exponential surface density profiles in the stellar population. With reasonable
assumptions about the scattering masses and sizes, the timescale for a rearrangement of
a disc from an initially flat profile to a exponential with several complete scale
lengths is about 1 Gyr. Soft scatterers like low-density clouds and big holes slow the
development of highly eccentric orbits compared to hard scatterers like dense clouds
(Fig. \ref{fig:ecc}). As a result, the stellar velocity dispersion increases more
slowly with soft scatterers even though the exponential disc develops quickly.
Evidently the exponential profile emerges with only moderate stellar orbit changes.

Our simulations isolated stellar scattering in order to avoid the confusion of collective
processes that occur in real galaxy discs. Similarly, \citet{se02} examined how spiral
waves drive radial migration from near the corotation resonance. This is a continuous
process for the duration of the wave and does not produce high eccentricity orbits. On
the other hand, the hard scattering by dense objects \citep{es13} is impulsive. Both
produce exponential discs.

We propose that an exponential disc that is moderately disrupted (e.g., by
interaction with a small companion) can restore itself by ISM scattering, making a
new exponential or piece-wise exponential on a Gyr timescale and erasing the most
obvious remnants of that disruption.  Relatively massive interstellar structures
are common in dIrr galaxies, possibly resulting from star formation feedback. They
seem to be the main driver of the continuous evolution toward stellar exponential
profiles, especially in dwarf galaxies.

We defined three phases of disc evolution (primarily in dwarf galaxies) under the
action of persistent weak scattering. The boundary between the first and second phases
was roughly defined by the morphological transition from initial transient waves to
smooth nascent exponential profiles. Phase 1 depends on the coldness of the initial
conditions and may not occur in real galaxies if they are continuously heated by
accretion and feedback. The boundary between the second and third phases was defined by
the kinematic transition between domination by ordered rotation versus dispersion.
Figure \ref{fig:foursnap} indicates that a relatively thick disc is preserved through
these transitions, and if the sources of scattering persist, may generate a discy,
early-type galaxy.

Another feature of our models is that the radial velocity dispersion is generally greater
than the vertical dispersion except in the outermost regions at late times. Part of this
larger radial dispersion is due to the fact that an early planar scattering event from a
circular orbit creates a radial velocity component and changes the angular momentum
simultaneously, while a vertical impulse does not have this double effect. More
importantly, since the stars are initially close to the mid-plane where the gas and star
formation are, the first scattering events are dominantly planar. Later scattering events
are less frequent because stellar orbits have been perturbed away from the scattering
clumps, and weaker because their orbits are no longer similar to those of the clumps. The
models suggest that the $z$ dispersion builds up over a long timescale, while the radial
dispersion appears quickly, and then evolves more slowly.

The thickening of the stellar disc because of scattering causes the evolution into
an exponential to slow down. This implies that new stars scatter somewhat independently
of old stars and make their own exponential profiles. A superposition of two exponential
profiles with different ages was demonstrated.

\section*{Acknowledgments}
We acknowledge use of NASA's Astrophysics Data System ADS. We also acknowledge discussions with J. Bird and S. Loebman, and many helpful suggestions from the anonymous referee.

\bibliographystyle{mn2e}

\bsp
\label{lastpage}
\end{document}